# La cosmología de la Divina Comedia

Alejandro Gangui

Instituto de Astronomía y Física del Espacio, CONICET,
Centro de Formación e Investigación en Enseñanza de las Ciencias y
Departamento de Física, Facultad de Ciencias Exactas y Naturales, UBA

> 'Oh!, turbia perspicacia;
> Oh!, observadores ciegos del cielo'.
> Tycho Brahe, *De Nova Stella*, 1573

En todas las épocas, la cosmología, la cultura y la civilización han estado ligadas en mayor o menor medida. La visión aristotélica de la física y del cosmos—una notable síntesis—fue, sin duda alguna, la más influyente en todos los ámbitos de la vida intelectual durante dos milenios. Sin embargo, la ciencia griega debió realizar un complicado periplo para llegar hasta nosotros. Durante la temprana y la alta Edad Media (aproximadamente entre los siglos V y XII), los conocimientos griegos se mantuvieron vivos en Europa occidental en forma muy fragmentaria. Pero en el Cercano Oriente se conservó y desarrolló aún más este precioso legado y fue desde allí que una 'nueva' cosmología reingresó en Europa. Las traducciones de las obras de Aristóteles y Ptolomeo, así como también tratados árabes y comentarios sobre los antiguos textos de ciencia griega, se difundieron en Occidente entre los años 1150 y 1300, sobre todo procedentes de la cultura islámica de España.

Pero la visión aristotélica no era propiedad exclusiva de los filósofos de la naturaleza; muchos escritores y poetas también se sintieron atraídos hacia ella. En todas las civilizaciones, la cosmología fue siempre un elemento clave de la cultura y, de una u otra manera, el movimiento de los cielos terminó impregnando la literatura de cada época. Entre los más notables poetas de la cultura occidental se encuentra Dante Alighieri (1265-1321) quien se hiciera célebre por su *Commedia*, escrita entre el año 1307 y la muerte de su autor y a la cual la crítica, a partir del siglo XVI, calificó de 'divina'. La *Divina Comedia* contribuyó fuertemente al desarrollo de la cultura popular de las ciudades-estado de Italia y colocó a la lengua italiana en un lugar preeminente dentro del marco europeo. Dante fue uno de los representantes del llamado *dolce stil novo*, movimiento literario que transformó la poesía popular amorosa en un arte refinado capaz de reflejar las influencias de corrientes filosóficas contemporáneas.

El cosmos de la *Divina Comedia* representa un modelo aristotélico simplificado en donde la Tierra permanecía quieta en el centro del universo y los astros eran transportados por esferas materiales cristalinas y transparentes. Según Aristóteles, la Tierra estaría rodeada por tres esferas sucesivas: de agua, de aire y de fuego. El conjunto de la Tierra y estas esferas elementales constituían el llamado mundo sublunar, más allá del cual se ubicaban las esferas de los planetas. Por fuera del conjunto concéntrico de esferas planetarias se ubicaba la esfera de las estrellas, a la que Dante otorga un movimiento casi imperceptible de oeste a este. Esto estaba de acuerdo con el movimiento de 'precesión de los equinoccios' que hoy sabemos vale unos 50,26 segundos de arco por año (o sea, unos 1,4 grados por siglo) y que fuera descubierto por Hiparco más de mil años antes de la época de Dante [ver Recuadro 1].

Más allá de la esfera de las estrellas se hallaba la esfera invisible que daba su movimiento a todas las esferas interiores, la morada del Primer Motor. Nuestro poeta florentino estaba lejos de ser un astrónomo principiante: Dante contaba con sólidos conocimientos de cosmología aristotélica y los estudiosos han encontrado en sus obras frecuentes y específicas referencias al cosmos tal como se lo entendía a fines del siglo XIII. Hay más de cien pasajes relacionados con la astronomía en la Divina Comedia, y cada uno de los cánticos de que se compone concluye con la palabra 'estrellas'.

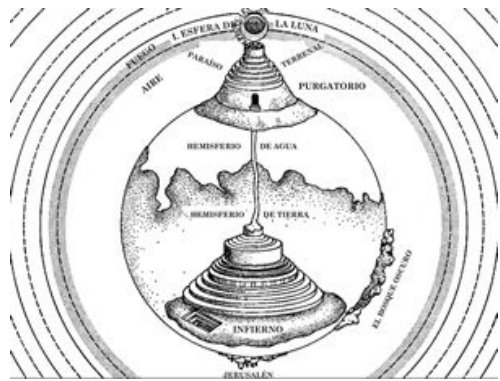

Figura 1: Una representación de la Tierra en la visión de la *Divina Comedia* de Dante Alighieri donde se muestra la ubicación—quizás en una forma demasiado 'realista'—de dos de los lugares de referencia, el Infierno y el Purgatorio. Surgidas de un común cataclismo, ambas formaciones tienen aspecto de cono. Los caminos de la pared cóncava interior del Infierno 'giran hacia la izquierda' mientras que las laderas convexas de la montaña del Purgatorio 'giran hacia la derecha'. Teniendo en cuenta que las posiciones del peregrino en uno y en otro están invertidas, el camino de Dante es siempre en forma de una espiral que progresa en la dirección del Empíreo y de la Divinidad.

Al narrar el viaje del poeta hacia el 'más allá', la *Divina Comedia* nos muestra la concepción del universo de la época. Este se divide en tres reinos, a los que corresponden las tres grandes partes del poema: 'Infierno', 'Purgatorio' y 'Paraíso'. El periplo comienza con Dante perdido en una selva oscura 'espiritual', esto es, un 'lugar' que no pertenece a la geografía física, sino que representa más bien un estado de ánimo de confusión mental y sentimental (contrariamente a las muchas representaciones pictóricas que ubican a esta selva en algún lugar de la superficie del globo terrestre) [ver Figura 1]. Guiado por el poeta romano Virgilio, Dante penetra a través de la boca del infierno y juntos descienden a las entrañas de la Tierra, recorriendo las pendientes del abismo del averno. En el infierno, los distintos niveles de castigos se ordenan en círculos de diámetro gradualmente decreciente, de modo tal que los peores pecadores se hallan a las mayores profundidades.

Al final de su descenso, Dante y su guía se encuentran con Lucifer. Este 'emperador del reino doloroso', que maneja el mundo terrestre, se halla quieto en su trono en el centro de la Tierra. Con la intención de dirigirse hacia el otro hemisferio, Virgilio—quien en este momento transporta a Dante—se aferra al flanco de Lucifer y desciende deslizándose por el cuerpo del gigante. Es así que, pasado un cierto punto, los viajeros quedan súbitamente dados vuelta y cabeza para abajo con respecto a Lucifer. Asombrado, Dante pide entonces a su guía que le explique lo sucedido, a lo que Virgilio responde (Infierno, canto XXXIV, 106-111):

*"... «Supones todavía*
 *hallarte tras el centro en que prendíme*
 *al gusano infernal* [Lucifer] *que horada al mundo.*
*Mientras yo descendí allí estuviste,*
 *cuando volví pasaste por el punto*
 *al cual tienden los pesos por doquiera»."*

*"... «Tu imagini ancora*
 *d'esser di là dal centro, ov' io mi presi*
 *al pel del vermo reo che 'l mondo fora.*
*Di là fosti cotanto quant' io scesi;*
 *quand' io mi volsi, tu passasti 'l punto*
 *al qual si traggon d'ogne parte i pesi»."*

Recordemos en este punto que Aristóteles afirma que todo elemento tiende a dirigirse hacia su lugar natural lo más rápidamente posible y por el camino más corto. El lugar natural de los objetos pesados (compuestos por el elemento 'tierra') es el centro de la Tierra y hacia allí caen en línea recta, tanto más rápido cuanto más pesados son. (Esta última, claro está, es una idea sobre la aceleración de los cuerpos que la física moderna desechará.) Como vemos, el lugar donde se ubica Lucifer representa para Dante el centro geométrico del universo y el lugar hacia donde todos los cuerpos pesados, compuestos de 'tierra', convergen—una clara herencia de las enseñanzas del gran filósofo griego.

Pero Lucifer no habría estado siempre en el centro de la Tierra, sino que habría caído allí, desde la cumbre de la Creación, como castigo junto con otros ángeles rebeldes. Esto lo explica nuevamente Virgilio en el último canto del Infierno (versos 121-126):

*"«Por esta parte él cayó del cielo;*
 *y la tierra que había en este sitio*
 *de pavor se veló bajo los mares,*
*luego emergió sobre nuestro hemisferio,*
 *y acaso por huirle dejó espacio*
 *a esa tierra que ves, y que se eleva»."*

*"«Da questa parte cadde giú dal cielo;*
 *e la terra, che pria di qua si sporse*
 *per paura di lui fe' del mar velo,*
*e venne a l'emisperio nostro; e forse*
 *per fuggir lui lació qui loco vòto*
 *quella ch'appar di qua, e sú ricorse»."*

Así, las tierras del sur retrocedieron por miedo a Lucifer, cubriéndose con las aguas a modo de un velo; se sumergieron en el océano y reemergieron en el hemisferio Norte formando la tierra firme conocida por los europeos. Dante introduce aquí una explicación claramente sobrenatural del antiguo interrogante de cómo el elemento pesado que es la tierra había logrado emerger por encima del elemento relativamente más liviano (el agua) para dar lugar a la tierra habitable.

Del centro de la Tierra, ambos personajes ascienden a través de un pasaje subterráneo hasta la costa de una isla del océano inexplorado del hemisferio sur. En esta isla se halla la montaña del Purgatorio, ubicada en dirección diametralmente opuesta a Jerusalén, y en cuya cima nuestros héroes encuentran el Jardín del Edén. Esta montaña 'que acá se ve elevada', se habría formado por un desplazamiento de tierra durante la caída de Lucifer, hecho cataclísmico que también habría generado la cavidad aproximadamente cónica del Infierno ('un vacío') por donde los viajeros habían descendido, tal como se relata en este primer cántico. Dado que el ápice de este cono llegaba hasta el centro de la Tierra, la montaña del Purgatorio—en última instancia el 'cono' de tierra que se desplazó y al hacerlo generó la cavidad del Infierno—debía ser increíblemente alta; de hecho, ya su tercera terraza estaba por encima de la atmósfera y su cima se ubicaba apenas por debajo de la esfera del fuego (la más externa de las esferas sublunares).

A partir de allí, Beatriz toma el relevo de Virgilio y conduce a Dante a través de la esfera del fuego y de las sucesivas esferas celestes del reino de los cielos, comenzando con la de la Luna. La forma de atravesar las esferas cristalinas utiliza la reflexión de la luz: en cada etapa Beatriz mira fijamente los engranajes celestiales mientras Dante observa la reflexión de estos en los ojos de su compañera. Cumplido este proceso, ambos son transportados inmediatamente al cielo siguiente.

En cada cielo los viajeros se encuentran con las almas de los bienaventurados. En la Luna, por ejemplo, encuentran a los inconstantes, aquellos que no cumplieron sus juramentos solemnes en la Tierra. Estos aparecen como meras imágenes difusas, reflexiones borrosas, tal como se manifestaba la Luna a los ojos de Dante. Mercurio alberga a los espíritus activos y a los líderes ilustres y en Venus se encuentran con los amantes famosos. Al Sol, que representaba la luz de la sabiduría, le corresponden los sabios, teólogos y filósofos. Así se sigue hasta Saturno, el más frío y alejado de los planetas, el séptimo cielo astronómico, donde se encuentran con los espíritus contemplativos.

Notemos que la morada de las almas en el 'más allá' era el Paraíso y que estas almas no residían en las esferas celestes, astronómicas, sino en el Empíreo, más allá del cielo de las estrellas. Los espíritus que se les aparecen a los peregrinos lo hacen para mostrarles la gloria gradualmente creciente de la que gozan, e indicarles sus antiguos temperamentos terrestres, los que a su vez habían sido influenciados por alguno de los siete astros mientras permanecían en la Tierra. Es de destacar también la importancia que Dante le otorga a la astronomía durante el desarrollo de este cántico: de los 33 cantos del Paraíso, 26 ocurren en el cielo astronómico por debajo de las esferas metafísicas (o teológicas) del Primer Motor y del Empíreo.

Dante y su compañera continúan entonces su travesía, dirigiéndose hacia las regiones más exteriores de este cosmos aristotélico cristianizado [ver Figura 2].

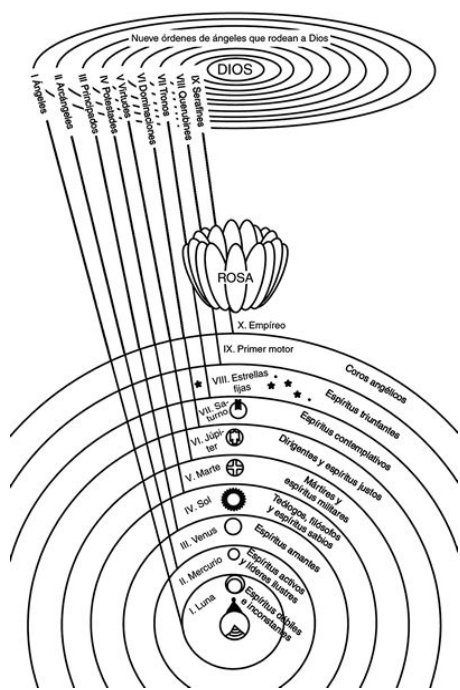

Figura 2: El cosmos de Dante agrega un mundo espiritual más allá del Primer Motor. A cada esfera del cielo astronómico le corresponde ahora una esfera de ángeles del mundo espiritual. La Tierra, centro del mundo físico, encuentra su contraparte celeste en Dios. Viajar alejándose de la Tierra ya no es más apartarse del 'centro' sino acercarse a lo divino.

Llegados a la esfera de las estrellas, el 'stellatum' u octavo cielo astronómico, Beatriz sugiere a Dante hacer una pausa y mirar el camino ya recorrido (Paraíso XXII, 124-129)

*"«Tan cerca estás de la salud excelsa»,
 dijo Beatriz, «que debes desde ahora
 tener los ojos claros y agudísimos;
pero antes de adentrarte más arriba,
 remira abajo, y cata cuánto mundo
 bajo tus pies he hecho que dejaras»;"*

*"«Tu se' sí presso a l'ultima salute»,
 cominciò Beatrice, «che tu dei
 aver le luci tue chiare e acute;
e però, prima che tu piú t'inlei,
 rimira in giú, e vedi quanto mondo
 sotto li piedi già esser ti fei»;"*

El peregrino hace caso y contempla desde las alturas los planetas, sus movimientos y distancias. Sonriendo ante la pequeñez y humildad de la Tierra comenta entonces (versos 133-135)

*"Con la mirada me volví hacia todas
las siete esferas, y tal vi a este globo
que sonreí al ver su vil semblante;"*

*"Col viso ritornai per tutte quante
le sette spere, e vidi questo globo
tal, ch'io sorrisi del suo vil semblante;"*

Seis horas más tarde, Dante volverá su vista nuevamente, esta vez para constatar cómo él y su compañera se han en efecto desplazado 90 grados, acompañando al cielo en su movimiento diurno.

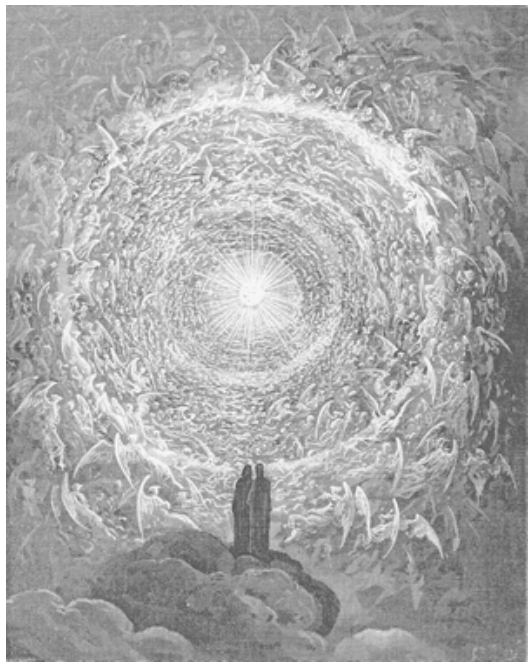

Figura 3: El Empíreo de Dante Alighieri, en un grabado de Gustave Doré (1832-1883), uno de los más famosos ilustradores de la Divina Comedia. Dante y Beatriz llegan al extremo del mundo físico y observan las esferas de ángeles del mundo espiritual: '...un Punto vi que allí irradiaba lumbre, / tan recia que los ojos que la enfocan / deben cerrarse por el fuerte brillo' (Paraíso, XXVIII, 16-18).

Llegados luego al 'Primum Mobile' (Primer Motor), Dante y Beatriz se hallan en el límite intangible entre lo natural y lo sobrenatural. Al mirar hacia arriba ven ahora un punto extraordinariamente pequeño y luminoso a la vez [ver Figura 3]. Alrededor de éste, nueve órbitas centelleantes representan las nueve órdenes de ángeles que rodean a Dios.

En el Paraíso XXVIII, 40-45, el poeta continúa

*"Mi dama, al verme así en tal cuidado*
 *fuerte suspenso, dijo: «De ese Punto*
 *depende el cielo y toda la natura.*
*Mira ese cerco que le está más próximo,*
 *y sabe que tan raudo es su moverse*
 *por el ardiente amor que le da impulso»."*

*"La donna mia, che mi vedea in cura*
 *forte sospeso, disse: «Da quel punto*
 *depende il cielo e tutta la natura.*
*Mira quel cerchio che piú li è congiunto,*
 *e sappi che 'l suo muovere è sí tosto*
 *per l'affocato amore ond' elli è punto»."*

Así, a diferencia del mundo físico aristotélico en el que los planetas más exteriores se desplazaban a mayor velocidad que los más cercanos al centro de giro (la Tierra) con el fin de completar la órbita diaria, en el mundo espiritual de la *Comedia* las órbitas eran tanto más 'divinas' (más rápidas) cuanto más interiores y cercanas a la luz central se encontraban. Entonces, Dante responde (versos 46-51):

*"Y a ella yo: «Si el mundo se asentase*
 *con el orden que veo en estas ruedas,*
 *saciado estaría yo con lo propuesto;*
*pero el mundo sensible nos permite*
 *ver esas ruedas tanto más divinas*
 *cuanto del centro se hallan más remotas»."*

*"E io a lei: «Se 'l mondo fose posto*
 *con l'ordine ch'io veggio in quelle rote,*
 *sazio m'avrebbe ciò che m'è proposto;*
*ma nel mondo sensibile si puote*
 *veder le volte tanto piú divine,*
 *quant' elle son dal centro piú remote»."*

Beatriz mostrará a Dante el motivo divino por el cual el mundo espiritual es así, no sin dejar de enumerar la entera jerarquía de ángeles y sus cualidades, según lo expuesto por el filósofo Pseudo-Dionisio [ver Recuadro 2], pues

*"«Y si tanta verdad secreta pudo*
 *revelar un mortal, tú no te asombres;*
 *pues quien aquí la vio pudo mostrársela,*
*junto a otras verdades de estos círculos»."*

*"«E se tanto secreto ver proferse*
 *mortale in terra, non voglio ch'ammiri;*

*ché chi 'l vide qua sú gliel discoperse*
*con altro assai del ver di questi giri»."*

Se revelaba así, en el final del canto XXVIII (versos 136-139), que si Pseudo-Dionisio conocía la verdadera distribución de las esferas de ángeles, era porque se lo había indicado el apóstol San Pablo, 'quien aquí la vio'—es decir, quien había estado en el cielo (ver Infierno II, 28)—.

Lo cierto es que con la integración del mundo espiritual al cosmos aristotélico, el genio de Dante proporcionaba una solución posible (dentro del marco de una obra literaria) a la incomodidad filosófico-religiosa que había caracterizado a los modelos cosmológicos de la Edad Media. Antes, elevarse hacia el cielo era hacer un viaje de alejamiento del 'centro', lo cual no era muy satisfactorio. Ahora, un viaje hacia lo más alto nos alejaba de la Tierra—en efecto—pero al mismo tiempo nos guiaba en la dirección de la divinidad, con etéreos coros angélicos mostrando el camino. Hacia mediados de la baja Edad Media, entonces, la *Divina Comedia* difunde un modelo de universo cristianizado en donde el mundo geocéntrico se muestra poseedor de un alma teocéntrica.

**Lecturas recomendadas**

D. Alighieri, *La Vita Nuova*, introd., coment. y glosario de Tommaso Casini, Sansoni, Florencia, 1885 (Biblioteca scolastica di classici italiani).

D. Alighieri, *La Vida Nueva*, Editorial El Aleph, Madrid, 1999.

D. Alighieri, *Commedia: Inferno*, ed. Emilio Pasquini y Antonio Quaglio, Garzanti, Milano, 1984 (2ª ed).

D. Alighieri, *Commedia: Purgatorio*, ed. Emilio Pasquini e Antonio Quaglio, Garzanti, Milano, 2002 (9ª ed).

D. Alighieri, *Commedia: Paradiso*, ed. Emilio Pasquini e Antonio Quaglio, Garzanti, Milano, 2002 (8ª ed).

D. Alighieri, *La Divina Comedia: Infierno.* Texto original italiano con trad., coment. y notas de Ángel J. Battistessa. Asociación Dante Alighieri, Buenos Aires, 2003.

D. Alighieri, *La Divina Comedia: Purgatorio.* Texto original italiano con trad., coment. y notas de Ángel J. Battistessa. Asociación Dante Alighieri, Buenos Aires, 2003.

RECUADRO 1:

Dante, en el comienzo de *La Vita Nuova* de 1293 (una recopilación de sus primeros poemas líricos), describe la edad de su amada Bice di Folco Portinari (Beatrice) a través de la fracción de grado angular que se ha movido la esfera de las estrellas en ese lapso de tiempo. En sus propias palabras:

> 'Luego de mi nacimiento, el luminoso cielo había vuelto ya nueve veces
> al mismo punto, en virtud de su movimiento giratorio, cuando apareció
> por vez primera ante mis ojos la gloriosa dama de mis pensamientos, a
> quien muchos llaman Beatriz, en la ignorancia de cuál era su nombre.
> Había transcurrido de su vida el tiempo que tarda el estrellado cielo
> en recorrer hacia Oriente la duodécima parte de su grado y, por
> tanto, aparecióseme ella casi empezando su noveno año [...]'.

Notemos aquí la diferencia entre 'el luminoso cielo', cuyo movimiento está dado por el Sol, y 'el estrellado cielo', cuyo lento movimiento está dado por las estrellas y es debido a la precesión de los equinoccios. ¿Qué es la precesión de los equinoccios? El eje de la Tierra no apunta siempre en la misma dirección con respecto a las estrellas lejanas: la Tierra rota sobre su eje, pero éste, bajo la influencia principal de la Luna, va cambiando lentamente su dirección como si fuera un trompo gigante. Debido a ello, la órbita de la Tierra alrededor del Sol (que determina el llamado plano de la eclíptica) no siempre cortará al plano ecuatorial de la Tierra en los mismos puntos (los equinoccios). Estos puntos se irán desplazando en el espacio—movimiento 'de precesión'—con un período que resulta ser de unos 25.765 años (el tiempo que el eje norte terrestre tarda en dar toda una vuelta alrededor de la constelación del Dragón). Ya mencionamos que el movimiento de precesión de los equinoccios representaba apenas unos 1,4 grados por siglo (o más precisamente, 1,396 grados por siglo), y es por ello que una vuelta de 360 grados será entonces completada en unos 25.765 años.

Ahora bien, Dante nos indica que el 'estrellado cielo' había recorrido un doceavo (1/12) de grado angular en el transcurso de la vida de Beatriz. Una simple 'regla de tres' nos permite entonces calcular la edad que debía tener su amada en ese primer encuentro:

    25.765 años x (1/12) / 360 = 6 años aproximadamente.

Pero Dante también menciona que Beatriz se le apareció 'casi empezando su noveno año', esto es, Beatriz apenas si llegaba a completar sus 8 años de edad. Conclusión, el verdadero valor era 8; él calcula 6, lo cual representa un 'error' (una subestimación) de apenas el 25%: ¡nada mal por tratarse de un poeta!

RECUADRO 2:

En realidad, Aristóteles postuló dos modelos diferentes de movimiento de las esferas celestes. En sus obras *Sobre el cielo* y la *Física* propuso la idea de que la esfera exterior o Primer Motor mueve a las esferas interiores por una acción mecánica o de fricción que se propaga. Pero dadas las obvias inconsistencias de esta explicación, en *Metafísica* XII, 7-8 la sustituye por otra según la cual en cada esfera habría una 'inteligencia' (sustancia inmaterial) que movería la esfera correspondiente 'por deseo'. Este segundo tipo de esquema es el que será favorecido por los intérpretes islámicos y hebreos medievales de Aristóteles, quienes identificaron las 'inteligencias' con los ángeles de la tradición bíblica—entre ellos, el médico y filósofo persa Avicena (siglo XI), pero también Algazel, Isaac y Maimónides. Esta identificación de la fuerza motriz de cada esfera con un ángel fue elaborada por Tomás de Aquino (siglo XIII), pero no fue aceptada por todos los escolásticos: Alberto Magno, el maestro de Tomás y una de las voces más autorizadas en filosofía de la naturaleza, la negaba, como también lo hacía el matemático y filósofo de la naturaleza Robert Kildwardby, manifestando la que era posición común en el Oxford de los siglos XIII y XIV.

Uno de las exposiciones más influyentes sobre la sistematización de las jerarquías angélicas fue debida a un autor neoplatónico del siglo VI AD que se conoció en la Edad Media con el nombre de pseudo Dionisio (debido a que se lo identificó equivocadamente con el personaje que, según el relato bíblico, San Pablo habría convertido durante su discurso en el ágora de Atenas). Entre las obras del pseudo Dionisio se encuentra una que se denomina *Sobre la jerarquía celeste*.

En el capítulo 6 de su obra *El Convivio*, Dante expone el esquema cósmico según el cual a cada esfera corresponde una de las jerarquías angélicas expuestas por el pseudo Dionisio. Es muy interesante advertir que en el capítulo 5, el poeta advierte sobre los dos modos diferentes de explicación que proporcionó Aristóteles en sus obras *Sobre el cielo* y *Metafísica*. Esto demuestra que sus textos literarios están asentados sobre un conocimiento reflexivo y profundo de la filosofía y astronomía de su época.